\documentclass[prc,noshowpacs,superscriptaddress,amsmath,amssymb,twocolumn]{revtex4-1}
\usepackage{epsfig}
\usepackage{verbatim}
\usepackage{multirow}
\usepackage{amsmath}
\usepackage{amssymb}
\usepackage{epstopdf}
\usepackage{color}
\usepackage[normalem]{ulem}
\begin{document}
	\title{
		Potential roots
of the deep sub-barrier heavy-ion fusion hindrance phenomenon}

	\author{P. W. Wen}
	\affiliation{Joint Institute for Nuclear Research, 141980 Dubna, Russia}
	\affiliation{China Institute of Atomic Energy,  102413  Beijing, China}

\author{C. J. Lin }
	\email{corresponding author: cjlin@ciae.ac.cn}
	\affiliation{China Institute of Atomic Energy,  102413  Beijing, China}
	\affiliation{Department of Physics, Guangxi Normal University, 541004 Guilin,  China}

	\author{R. G. Nazmitdinov}
	\email{corresponding author: rashid@theor.jinr.ru}
	\affiliation{Joint Institute for Nuclear Research, 141980 Dubna, Russia}
	\affiliation{Dubna State University, 141982 Dubna, Russia}
	
\author{ S. I. Vinitsky}
	\email{corresponding author: vinitsky@theor.jinr.ru}
	\affiliation{Joint Institute for Nuclear Research, 141980 Dubna, Russia}
	\affiliation{Peoples' Friendship University of Russia (RUDN University), 117198  Moscow, Russia}

	\author{O. Chuluunbaatar}
	\affiliation{Joint Institute for Nuclear Research, 141980 Dubna, Russia}
	\affiliation{{Institute of Mathematics and Digital Technologies, Mongolian Academy of Sciences, 13330 Ulaanbaatar, Mongolia} }

	\author{A. A. Gusev}
	\affiliation{Joint Institute for Nuclear Research, 141980 Dubna, Russia}

	\author{A. K. Nasirov}
	\affiliation{Joint Institute for Nuclear Research, 141980 Dubna, Russia}
	\affiliation{Institute of Nuclear Physics, Ulugbek, 100214, Tashkent, Uzbekistan}
		
	\author{H. M. Jia}
	\affiliation{China Institute of Atomic Energy,  102413  Beijing, China}
	
	\author{A. G\'o\'zd\'z}
	\affiliation{Institute of Physics, University of M. Curie{-}Sk{\l}odowska, 520031 Lublin, Poland}
	
	\date{\today}
\begin{abstract}
We analyse the origin of the unexpected deep sub-barrier heavy-ion fusion hindrance
in  ${}^{64}$Ni+${}^{100}$Mo and  ${}^{28}$Si+${}^{64}$Ni recations.
Our analysis is based on the improved coupled-channels  approach, implemented  by means
of the finite element method. With the aid of the Woods-Saxon potential the experimental cross
sections and the $S$-factors of these reactions   are remarkably well reproduced.
We found that the account on the non-diagonal matrix elements of the coupling matrix,
traditionally neglected in the conventional  coupled-channels approaches
in setting the left boundary conditions inside
the potential pocket, and its minimal value are crucially important for
the interpretation experimental data.
Within our approach we found a good agreement with
the experimental data for the S-factor of the fusion reaction ${}^{12}$C+${}^{12}$C,
which  has no a pronounced maximum for this system.
\end{abstract}
\pacs{25.70.Jj, 25.70.Hi, 24.10.Eq}
\maketitle
	
Various stages of astrophysical nucleogenesis, the synthesis of superheavy nuclei, and, consequently, effective mechanisms of nucleus-nucleus interaction ~\cite{montagnoli17,back14,linbook,yang17,jia16,nasirov19},
require a deep understanding of the near barrier heavy-ion fusion reaction.
Since 2002  precise measurements  have been available to probe  the effects
of the interaction potential at the deep sub-barrier energies~\cite{jiang02}.
However, at these energies the fusion cross sections fall off much steeper than the conventional coupled-channels (CC) calculations  predict. This unexpected fusion hindrance phenomenon is generally accompanied by the maximum of the astrophysical $S$-factor. Due to its notable influence on the astrophysical nuclear reaction processes for fusion reactions like
$^{12}$C+$^{12}$C ~\cite{jiang18-2,tumino2018,tan2020,fruet2020,Liyj2020,beck2020},
this problem is a subject of extensive studies in the past years~\cite{jiang02,jiang05,dasgupta07,jiang16-2}.

Since the discovery of the fusion hindrance phenomenon, the consensus is that the conventional CC calculations based on a Woods-Saxon (WS) potential are insufficient to reproduce the experimental data~\cite{jiang04,montagnoli17,back14}. As a result, numerous theoretical attempts have been
developed to tackle this problem. Among them are such as the hybrid of different nuclear potentials~\cite{misicu06,hagino18}, the extending of the CC framework to the adiabatic states~\cite{ichikawa09}, the quantum diffusion approach~\cite{sargsyan20}, the density-constrained frozen Hartree-Fock method~\cite{simenel17},
to name just a few.

In fact, within the conventional CC approach we can distinguish between
two directions of the theoretical explanations for the hindrance mechanisms.
The first one is based on the sudden approximation for a hybrid  of different potentials.
In particular, a gentle overlap of the reacting nuclei is considered
due to the saturation properties of nuclear matter~\cite{misicu06,misicu07,esbensen11}.
One attempts to explain the steep falloff of fusion cross sections
by using the double-folding potential with M3Y forces supplemented by a repulsive core.
Another approach analyses the hindrance phenomenon  by fitting the fusion excitation function with two separate WS potentials~\cite{hagino18}. On different sides of the threshold energy where the maximum of the $S$-factor locates,  the model includes the potentials that produce different logarithmic slope of the excitation functions.
The second direction is based on the adiabatic approximation~\cite{ichikawa07,ichikawa09}. On the top of the conventional CC method, an extra one-dimensional adiabatic potential barrier is assumed after the reacting nuclei contacts with each other, considering the formation of the composite system.
 Thus, the mechanism of the deep sub-barrier hindrance is still debatable.
In spite of numerous attempts it remains to be a real challenge for nuclear reaction theory.	
The main goal of the present paper is to explain
the unexpected deep sub-barrier heavy-ion fusion hindrance,
providing the principle of bounding solutions for the
scattering problem of two colliding nuclei, developed in Refs.~\cite{wen2020-1,vinitsky2020-1}.

We consider the collision between two nuclei, which relative motion is coupled to nuclear intrinsic motion.
	The potential between the projectile and the target
	contains the  Coulomb potential  and  nuclear potential,
	chosen in the WS form
	$V^{(0)}_N(r)=-{V_0}/{(1+\exp[(r-R_0)/a_0])}\,.$
	Here, the parameters $V_0$, $R_0$, and $a_0$ are  the potential depth, potential
	radius, and diffuseness, respectively{; $r$ is the distance between
		the mass centers of the two interacting nuclei}.

	\begin{figure*}
		\begin{center}
			\includegraphics[width=15cm,angle=0]{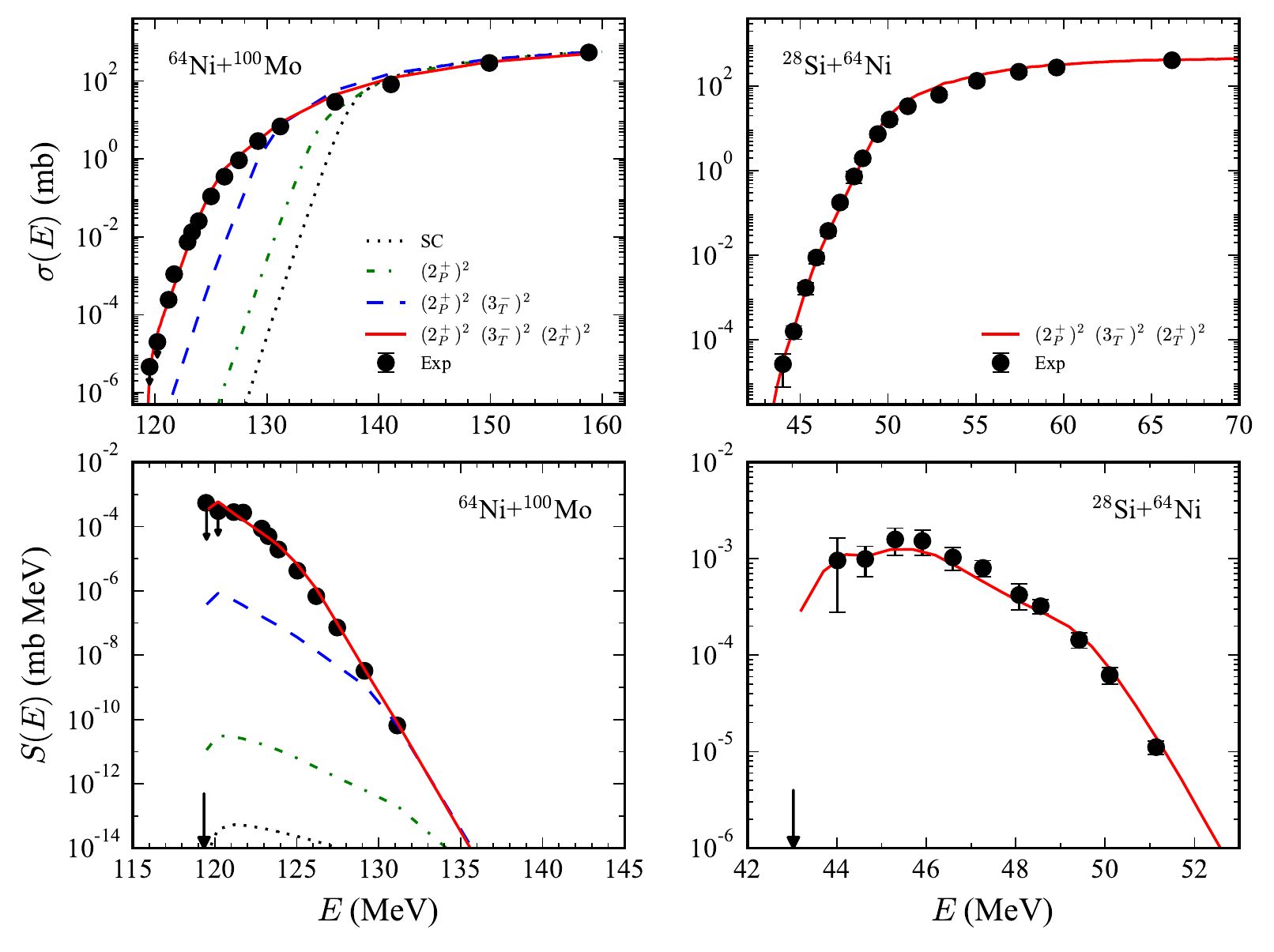}
			\caption{
(Color online) The fusion cross sections $\sigma(E)$ and the astrophysical $S(E)$-
factor for  ${}^{64}$Ni+${}^{100}$Mo and   ${}^{28}$Si+${}^{64}$Ni reaction systems.
Different curves denote the CC calculations with different sets of  collective vibrations,
indicated by the legends; the SC means a single-channel calculations (without coupling).
The experimental data (solid circles) are taken from Ref.\,\cite{jiang05} and Ref.\,\cite{jiang16-2},
respectively. The black arrows  indicate  the potential pocket minimum $V_{\rm P}$.}
			\label{cs}
		\end{center}
	\end{figure*}
	
	Following Refs.\,\cite{hagino99,hagino12}), in our approach the nuclear coupling Hamiltonian is generated by changing the target radius in the potential to the dynamical operator $R_0+\hat{O}$, which is related to collective vibrations. The solution of the CC equations between values $r_{\rm max}$ and  $r_{\rm min}$ is found under the incoming wave boundary conditions (IWBC), i.e., it is assumed a strong absorption inside the potential pocket. The right boundary point $r_{\rm max}$ is usually set at a large enough distance where the interaction is weak, and the off-diagonal elements of the coupling matrix tend to be zero. The  $r_{\rm min}$ is determined at the minimum of the  potential pocket $V_P$.
	
At the left boundary $r = r_{ \rm min}$, the open
left exit channel wave functions are usually taken as the plane wave
${\psi}_{nn_o}^{(\ell)}(r)=\exp\left(-i k_n( r_{ \rm min}) r\right) {T}^{(\ell)}_{n n_o}$,
where ${T}^{(\ell)}_{n n_o}$ is the tunneling amplitude.
The definition of $k_{n}(r_{\min})=\sqrt{\frac{2\mu}{\hbar^2}E-W^{(\ell)}_{nn}(r_{\min})}>0$,
involves only diagonal elements of the coupling matrix,
assuming that the off-diagonal matrix elements tend to be zero
(e.g., Refs.\,\cite{hagino99,hagino12}).
We recall  that  $ W^{(\ell)}_{nm}(r)=\frac{2\mu}{\hbar^2}
\left[V^{(\ell)}(r)\delta_{nm}+V_{nm}(r)\right]$
at $r > r_{\min}$;
and  the constant  valued matrix  $W_{nm}=W^{(\ell)}_{nm}(r_{min})$ at $r\leq r_{\min}$.
Here, $V^{(\ell)}(r)=\frac{Z_P Z_Te^2}{r}+V_N^{(0)}(r)+ \frac{\hbar^2\ell(\ell+1)}{2\mu r^2}+\epsilon_n$
is the potential energy without the coupling. Further,
$V_{nm}(r)$ are elements of the coupling matrix,
$\mu$ is the reduced mass, $Z_P$ and $Z_T$ are the Coulomb charges of projectile and target  ions,
$\ell$ is the orbital angular momentum, and $\epsilon_n$ is the excitation energy
of the $n$-th entrance channel or the entrance threshold energy $E=\epsilon_n$  at $n=1,...,N$.

It is important to stress that at $r_{\rm min}$, the distance between two nuclei becomes so small,
that the off-diagonal matrix elements $W^{(\ell)}_{nn'}(r_{\rm min})$ are usually not zero.
As addressed in Refs.\,\cite{zagrebaev04,samarin04}, there can be sudden
noncontinuous changes at the left boundary conditions, and this will distort the total wave
function inside the barrier.

To treat properly this problem, at the left boundary we adopt
the linear transformation method~\cite{wen2020-1,vinitsky2020-1}.
Namely, at $r \le r_{\min}$, when the off-diagonal matrix elements have been taken into account,
the  modified solutions of the CC equations ${\tilde \psi_{nn_{o}}^{(\ell)}(r)}$ consist of
the linear independent solutions $\phi_{nm}^{(\ell)}(r)$, i.e.,
${{\tilde \psi}_{nn_o}^{(\ell)}}(r)=
\sum_{m=1}^{M_o}\phi_{nm}^{(\ell)}(r){\tilde T}_{mn_o}^{(\ell)}$.
In this case the linear independent matrix solution
	can be obtained by considering the transformation
	$    \phi_{nm}(r)=A_{nm}y_m(r),\label{lineareq}$
	where $y_m(r)$ are solutions of the uncoupled equations
	\begin{equation}
	y''_m(r)+ K_m^2 y_m(r)=0,  K_m^2=(2\mu/{\hbar^2}) E-\tilde{W}_{mm}\nonumber\,.
	\end{equation}
Here, ${\bf A}$ and $\tilde {\bf W}$ are the matrix of eigenvectors and
the diagonal matrix of eigenvalues of the eigenvalue problem, respectively.
In short, we diagonalize the coupling matrix
${\bf A}^{-1}{\bf W}{\bf A}=\tilde {\bf W}$.
For the open channels with $K_m^2>0$, 	$y_m(r)={\exp(-i K_m r)}/{\sqrt{K_m}}$  at $m=1,...,M_o\leq N$.
The partial fusion probability
$P_{n_o}^{(\ell)}=\sum_{m=1}^{M_o}|{\tilde T}_{mn_o}^{(\ell)}|^2\neq \sum_{m=1}^{M_o^{\prime}}|{T}_{mn_o}^{(\ell)}|^2$
is given by summing over all open exit channels at $r \le r_{\rm min}$.
We stress that the number of open channels after and before the diagonalization of the coupling matrix
will be different.
At $r\geq r_{\max}$ the asymptotic solutions are given in terms of the normalized  Coulomb functions
with the wave number $k_{n}=\sqrt{\frac{2\mu}{\hbar^2}(E-\epsilon_{n})}>0$ in
the entrance $n_o$ and the right exit $n$ open channels $n_{o},n=1,...,N_{o}\le N$ and
the reflection amplitudes ${\tilde R}_{nn_o}^{(\ell)}$ \cite{vinitsky2020-1}.

	
To illuminate all cons and pros of our approach	we reexamine
${}^{64}$Ni+${}^{100}$Mo, and   ${}^{28}$Si+${}^{64}$Ni reactions.
To this aim we analyse the fusion cross sections
$\sigma=\sigma_{\rm fus}(E)=\sum_{\ell=0}^{\ell_{max}}\sigma_{\ell}(E)=
\frac{\pi}{k_{n_{o}}^2}\sum_{\ell=0}^{\ell_{max}}(2\ell+1)P_{n_{o}}^{(\ell)}(E)$ and
the astrophysical $S$ factor (see Fig.\,\ref{cs}).
Our results  have been obtained with the aid
of the KANTBP code, developed by means of the finite element method (see for details Refs.~\cite{chuluunbaatar07,chuluunbaatar08,gusev14-2,gusev15-2,Chuluunbaatar2020})
for the set $N=1+N_{coupl}$   of CC equations with the improved IWBC
from the corresponding ground states $|n_{o}-1=0>$.
The number of all channels $N_{coupl}$ depends of the chosen reaction.
It is noteworthy that, using the program KANTBP, the sum of
tunneling  and reflection probabilities
$\sum_{m=1}^{M_o}|{\tilde T}_{mn_o}^{(\ell)}|^2+\sum_{n=1}^{N_o}|{\tilde R}_{nn_o}^{(\ell)}|^2-1\simeq 10^{-10}$.
It is the stringiest test of the validity of our calculations.

The coupling radius parameter for the collective vibrations is set to be 1.2 fm for
all the cases in this study \cite{wen2020-1}. Note, that the value of $\ell_{max}$
is restricted by {\it the constraint} on  the incident  energy values $E$ in the entrance channel:
$E=V^{(\ell)}(r_{min})$,  where $V^{(\ell)}(r_{min})$ is the potential minimum,
and $\ell=0,...,\ell_{max}$.
The  values of $\eta_0$ used for scaling the astrophysical factor
$S(E)=E\sigma_{\rm fus}(E) \exp(2\pi(\eta-\eta_0))$  for the above three reactions
are  105.74, 75.23, and 41.25, respectively; $\eta$ is the Sommerfeld parameter.
The adopted structure properties (including excitation energies,
deformation parameters for the collective state) are taken from Refs.\,\cite{raman01,kibedi02}.
The potential parameters in this study are obtained by  fitting the experimental fusion data
at the whole energy region with the CC calculations and the simple WS potential (see Table \ref{fit}).

	\begin{table}
		\caption{ Woods-Saxon potential parameters $V_0$ (MeV),  $a_0$ (fm), $R_0$ (fm) for ${}^{64}$Ni+${}^{100}$Mo, and ${}^{28}$Si+${}^{64}$Ni reaction systems. The potential barrier $V_{\rm B}$ and the minimum of the potential pocket $V_{\rm P}$  are also listed. }
		\begin{ruledtabular}
			\begin{tabular}{lrrrr}  \label{fit}
				&  ${}^{64}$Ni+${}^{100}$Mo  &        ${}^{28}$Si+${}^{64}$Ni      \\
				\hline
				$V_0$    (MeV)        & 79.938  & 62.707       \\
				$a_0$  (fm)           &0.686  &  1.014      \\
				$R_0$ (fm)             & 10.190  &  7.354            \\
				$V_{\rm B}$ (MeV)       & 136.993  &  52.697           \\
				$V_{\rm P}$ (MeV)       & 119.344  &  43.027          \\
			\end{tabular}
		\end{ruledtabular}
	\end{table}

There is a remarkable agreement between our calculations  and available experimental data for  the fusion cross sections and $S$-factors (see Fig.\,\ref{cs}).   Note, that all $S$-factors
of these reactions have  maxima.
To explore the general reason for the hindrance and the maximum of the $S$-factor, we consider also the results for different combinations of the collective vibrations for ${}^{64}$Ni+${}^{100}$Mo (see Fig.\,\ref{cs}).
All calculations for various number of coupled channels
demonstrate as well the maximum for $S$-factor, including
the single-channel  case (SC) when all $V_{nm}(r)\equiv 0$ (i.e. without couplings).
We observe that the energies, where the hindrance and the maximum of $S$-factor take place,
 are close to  the potential pocket minimum $V_{\rm P}$ for different couplings.

\begin{figure}
		\begin{center}
			\includegraphics[width=7 cm,angle=0]{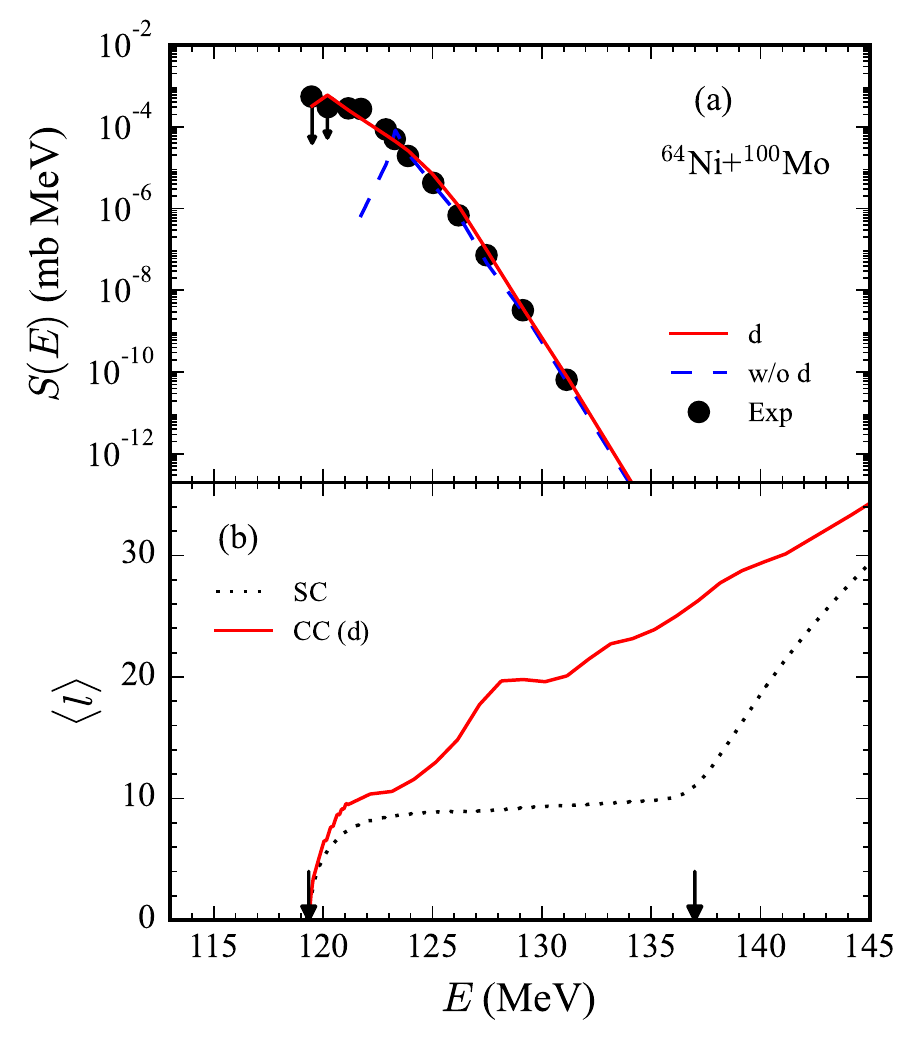}
			\caption{(Color online)  The astrophysical $S(E)$-factor  and the mean
            orbital momentum  $\langle \ell \rangle$ for the reaction ${}^{64}$Ni+${}^{100}$Mo.
            Panel (a): the results of
            the CC calculations with the diagonalization (solid line) and without the diagonalization
            (dashed line) are compared with the experimental data (full circle) [see text for details].
            Panel (b): the mean orbital momentum  $\langle l \rangle$  for the a single channel (without coupling) (dotted line) and for the full coupling (solid line) with the diagonalization procedure.
			 The arrows indicate the position of the potential barrier $V_{\rm B}$ and
              the  pocket minimum energy $V_{\rm P}$.}
			\label{nl}
		\end{center}
	\end{figure}

Although  the importance of the potential pocket minimum  in
the CC calculations  was already noticed in Refs.~\cite{misicu06,misicu07,montagnoli13},
the agreement between experimental data and the results of calculations was not
reached. In these studies the repulsive core inside the shallow potential pocket was
suggested as one of the  reasons for the hindrance phenomenon.
To find out why the hindrance and the maximum of $S$-factor happen always near $V_{\rm P}$,
we compare the CC results  without and with the diagonalization for
${}^{64}$Ni+${}^{100}$Mo [see Fig.\,\ref{nl}~(a)]. It appears that the correct treatment of the
left boundary is one of the decisive factors, that allows to reach a good
agreement with the experiment, using the simple WS potential.

To gain  further insight into the details of the hindrance phenomenon, we compare
the  mean angular momentum
 $\langle \ell\rangle =\sum_{\ell=0}^{\ell_{max}}\ell \sigma_{\ell}(E)/\sigma(E)$ (see also
 Ref.~\cite{hagino97-2})
  for the complete calculations and without coupling   [see Fig.\,\ref{nl} (b)].
 Note, that  when  $E\rightarrow V_{\rm P}$,  the  $\langle \ell \rangle$ decreases
  to zero quickly if there is the constraint on the energy for both cases.
 In this case, the energy is too small, and only the $s$-wave partial contribution
 determines the cross section.
It seems, that to obtain a good agreement with the experimental data,
the constraint is important as well [compare Figs.\,\ref{nl}~(a),~(b)].
At   $E>V_{\rm P}$ there are many coupled channels at the complete calculations, and,
consequently, the barrier has a certain kind of distribution, which obscures the barrier position.

	\begin{figure}
		\begin{center}
			\includegraphics[width=7cm,angle=0]{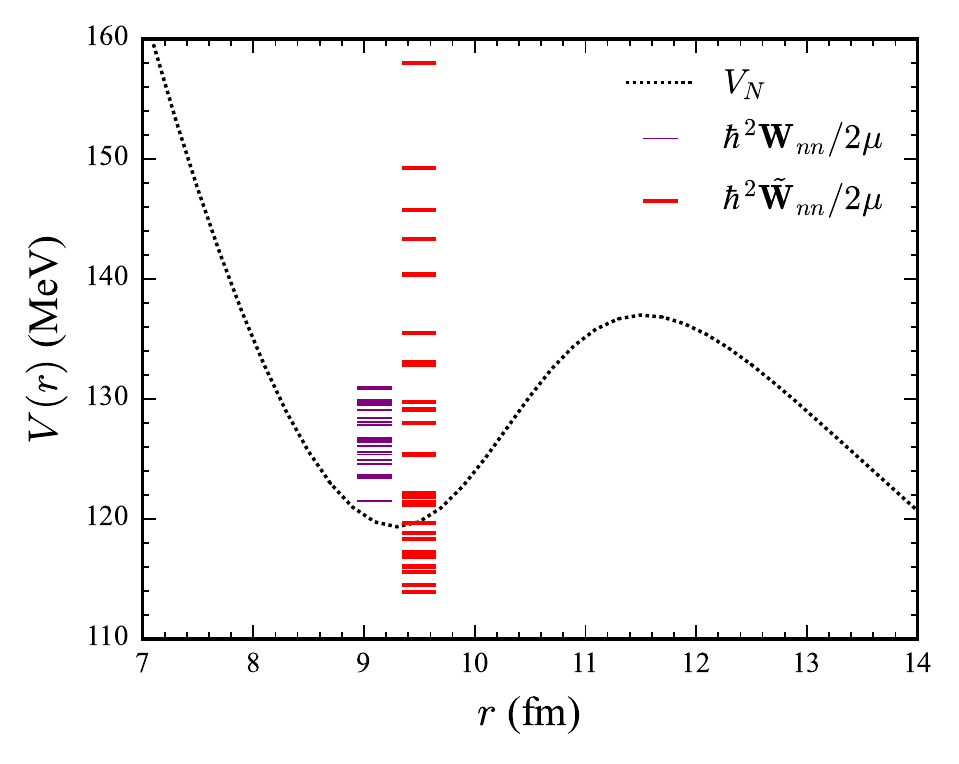}
			\caption{(Color online) The potential $V(r)$ (dotted line),
             diagonal matrix elements of the coupled matrix $\hbar^2 {\bf W}_{nn}(r_{\min})/2\mu$
             (thin solid lines) and eigenvalues $\hbar^2 {\bf\tilde{W}}_{nn}(r_{\min})/2\mu$
             (thick solid lines)   for the case of ${}^{64}$Ni+${}^{100}$Mo. See text for details. }
			\label{potential}
		\end{center}
	\end{figure}

To elucidate further the  basic mechanism of the hindrance factor in our calculations,
we compare the potential energy $V(r)$ (without coupling),
the diagonal elements $\hbar^2  {\bf W}_{\rm nn} / 2\mu$ of the coupled matrix,
and the threshold energies $ \hbar^2 \tilde {\bf W}_{\rm nn} / 2\mu$
at the left boundary (see Fig.\,\ref{potential}).
As it is seen, the threshold  energies  $ \hbar^2 \tilde {\bf W}_{\rm nn} / 2\mu$
spread much wider than the
diagonal elements $\hbar^2  {\bf W}_{\rm nn} / 2\mu$ of the coupled matrix.
Especially, the minimum  threshold energy of $\hbar^2\tilde {\bf W}_{\rm 11} / 2\mu$ is
obviously much lower  $V_P$.
In other words, in contrast to the conventional CC calculations, in our approach
the number of open channels is much larger.

It appears that the experimental fusion cross section can be reproduced well only
under certain physical couplings. For these reactions, the entanglement between
the states at the left boundary is changed through the diagonalization procedure.
On the other hand, in the CC calculations, when the incident energy
$E<V_{\rm P}=V^{(\ell=0)}(r_{min})$, the tunneling is absent.
It is due the fact that the ingoing flux will be zero~\cite{hagino99,misicu07,montagnoli13}.
Thus, when the incident energy gradually approaches the bottom of the potential pocket
minimum, the fusion hindrance occurs naturally.

	\begin{figure}
		\begin{center}
			\includegraphics[width=7cm,angle=0]{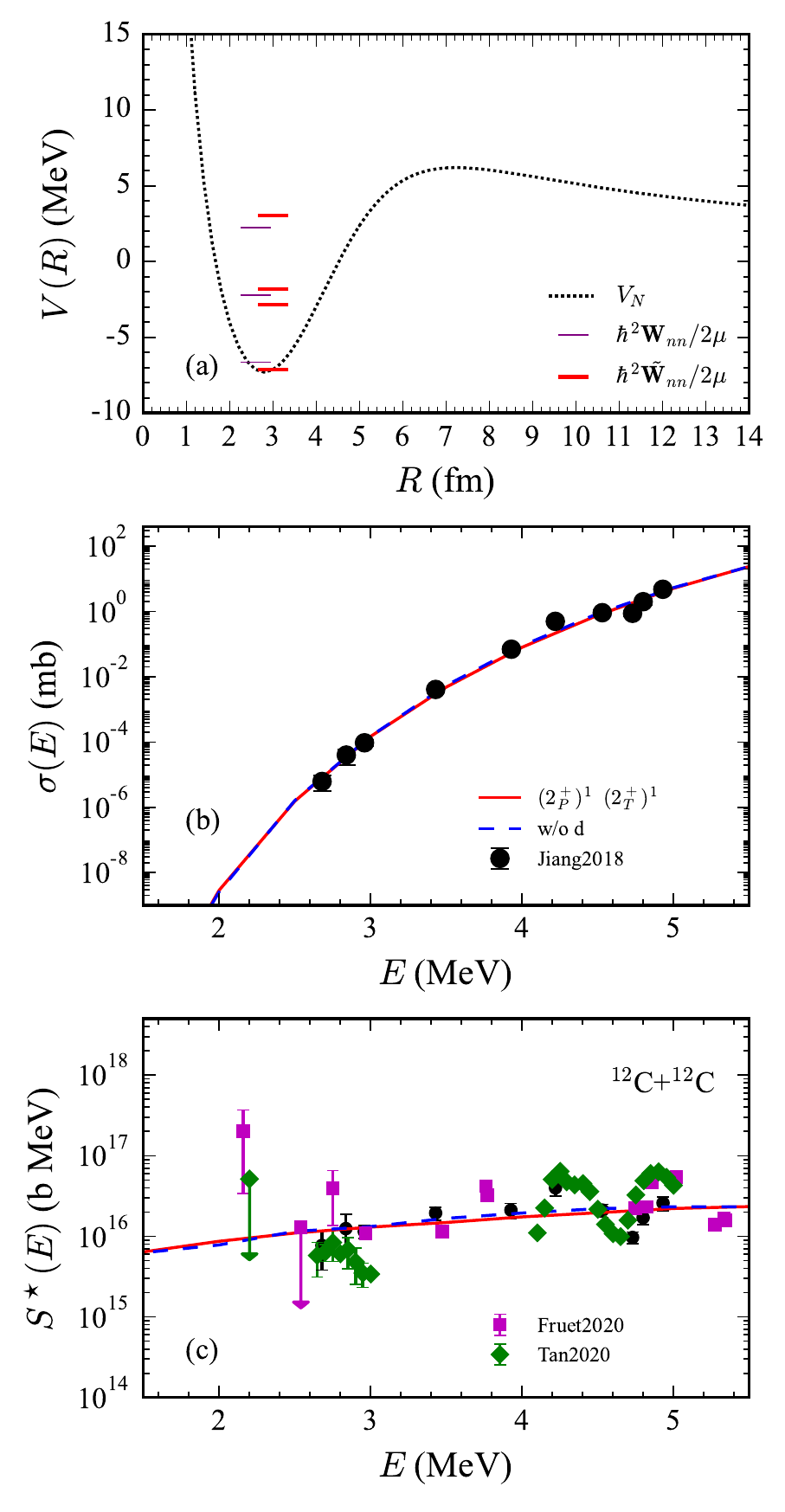}
			\caption{(Color online) (a) Similar to Fig.\ref{potential} for the case of
				$^{12}$C+$^{12}$C.
				(b) The results for the fusion cross sections $\sigma(E)$
                    with the diagonalization (solid line) and
                   without the diagonalization (dashed line).
				(c) The results for the $S^\star (E)$-factor with the diagonalization (solid line) and
                   without the diagonalization (dashed line).
				The experimental data labeled as Jiang2018, Tan2020, Fruet2020, are
                 taken from Refs.\,\cite{jiang18-2}, Ref.\,\cite{tan2020}, Ref.\,\cite{fruet2020}, respectively.
				All results are obtained  with the indicated collective vibrations.
			}
			\label{12C}
		\end{center}
	\end{figure}
	
Let us turn to the most important fusion reaction
$^{12}$C+$^{12}$C in nuclear astrophysics within our aprroach. We recall that the carbon fusion plays a significant role in the burning of the massive stars, ignition of the type Ia supernovae explosion, superbursts of binary systems or neutron stars ~\cite{back14}.
It remains an open problem whether the fusion hindrance does, indeed, occur in this reaction,
which  is closely related to the astrophysical reaction rate~\cite{jiang18-2,beck2020}.

As above, we fit the experimental fusion cross section~\cite{jiang18-2}  with the aid of the  WS potential:
$V_0=34.252$ MeV, $R_0=3.865$ fm, and $a_0=0.952$ fm. In addition, we consider the quadrupole excitations.
In this case the fusion cross section is defined as
$\sigma(E)=2\sum_{\ell={\rm even}}^{\ell_{max}}\sigma_{\ell}(E)$.
For the carbon fusion, we adopt the commonly used definition
$S^\star(E) = \sigma(E) \cdot E \cdot \exp(87.21/\sqrt{E}+0.46 E)$~\cite{Liyj2020,tan2020}.
The results of calculations with and without the diagonalization demonstrate a good agreement
with experimental fusion data ~\cite{jiang18-2,tan2020}.
In contrast to the results shown in Fig.\ref{cs}, where the $S(E)$factor drops
at  low energy tail region evidently, the  $S^\star(E)$-factor  evolve smoothly with the decreasing energy.
It should be noted that the trend of the $S^\star(E)$-factor over the energy $E$
is different from that for the $S(E)$-factor.
At low energy   region,  $S^\star(E)$ is lower than that of $S(E)$. And for the $S^\star(E)$-factor,
the calculations indicate that there are no clear decrease and  the maximum for this system at low energies.
Note, that our results are similar to those of the CC theory with M3Y+repulsive core potential in Ref.\,\cite{esbensen11}.
	
In our calculations, based on the WS potential, the reason for the steady trend
can be traced from Fig.~\ref{12C}a.
The threshold energies after the diagonalization change modestly in the comparison with those without the diagonalization. The bottom of the potential pocket is about -7 MeV, which is far
from the incident energy region of interest (about 1.5-3 MeV).
Therefore, the hindrance feature is not so obvious as that seen in Fig.~\ref{cs}.
In the former case the $S^\star$ factor changes slowly below the potential barrier.

Surprisingly, we find that our results are supported by the empirical trends,
discussed
for the hindrance factor in Ref.~\cite{jiang06}.
Indeed, our results for the medium nuclei manifest the hindrance factor for system with
$Z_1Z_2\sqrt{\mu}\geq 2000$. While for the lightest system with $Z_1Z_2\sqrt{\mu}\leq 200$ the logarithmic
slopes of the $S^\star(E)$ factor exhibits resistance to the increasing tendency with the energy.
In our approach the variation of the coupling strength of the left exit
channels is the basic mechanism, responsible for the observed phenomenon.
The coupling strength is  much stronger for the medium and heavy system, while it is much weaker
for the lightest systems. In other words, the degree of the strength controls the number of
open channels, contributing to the reaction.

	In summary, the deep sub-barrier heavy-ion fusion hindrance phenomenon
and the behavior of the astrophysical $S$-factor  for   ${}^{64}$Ni+${}^{100}$Mo,
and  ${}^{28}$Si+${}^{64}$Ni reactions are analysed by solving the CC equations with the
improved IWBC. This approach has been developed in Refs.\,\cite{wen2020-1,vinitsky2020-1} and based on the finite element method KANTBP ~\cite{chuluunbaatar07,chuluunbaatar08,gusev14-2,gusev15-2,Chuluunbaatar2020}).
The obtained results  reproduce remarkably well the experimental data with the aid of the simple WS potential.
It is found that the calculated $S$-factors with different kinds of collective vibrations have  maxima
for the considered reactions. The knowledge of the potential minimum energy $V_{\rm P}$ and
the improved IWBC are  crucially important for the correct interpretation of the fusion cross section in the
conventional CC calculations.
The general trend of the  directly measured fusion data for the
reaction $^{12}$C+$^{12}$C ~\cite{jiang18-2,tan2020} is  described as well.
It is found that the $S^\star$ factor drops gently at energies under the Coulomb barrier,
and the results show no a pronounced maximum of  the $S^\star$-factor for this system. We hope that further
experiments at low energies could finally reveal  "the mystery of the hindrance phenomenon"
as a function of the mass number. From our point of view it is simply determined by the
number of coupled channels at the correct treatment of the left boundary conditions for
different combinations of the colliding nuclei.

	\vskip5mm
	
	\section*{ACKNOWLEDGEMENTS}
The work of P.W.W., C.J.L., and H.M.J. is supported by the National Natural Science Foundation of China (Grants Nos. 11635015, {11805120,} 11635003, 11805280, {11811530071, U1867212} and U1732145), the National Key R\&D Program of China (Contract No. 2018YFA0404404), and the Continuous Basic Scientific Research Project (No.WDJC-2019-13).
The present research benefited from computational resources of the HybriLIT heterogeneous platform of the JINR. This work was partially supported by the Polish--French COPIN collaboration of the project 04--113, the Bogoliubov--Infeld and the Hulubei--Meshcheryakov JINR	programs, the grant RFBR and MECSS 20--51--44001,the grant RFBR 17-52-45037,
 the RUDN University Program 5--100	and grant of Plenipotentiary of the Republic of Kazakhstan in JINR.


%

\end{document}